\documentclass[useAMS,usenatbib]{mn2e}

\usepackage{amssymb}
\usepackage{aas_macros}
\usepackage{natbib}
\usepackage{times}
\usepackage[dvips]{graphicx}

\title[The effects of pre-virialization and photoheating on the baryon fraction of $\Lambda$CDM haloes]{The baryon fraction of $\Lambda$CDM haloes}  \author[R.~A.~Crain et al.]{
\parbox[h]{160mm}{Robert A. Crain$^1$\thanks{E-mail:
r.a.crain@durham.ac.uk}, Vincent R. Eke$^1$, Carlos S. Frenk$^1$, Adrian Jenkins$^1$,\\ Ian G. McCarthy$^1$, Julio F. Navarro$^2$ \& Frazer R. Pearce$^3$}\vspace{6pt}\\
$^1$Institute for Computational Cosmology, Department of Physics, University of Durham, South Road, Durham, DH1 3LE, UK\\
$^2$Department of Physics \& Astronomy, University of Victoria, Victoria, British Columbia, V8P 1A1, Canada\\
$^3$School of Physics \& Astronomy, University of Nottingham, University Park, Nottingham, NG7 2RD, UK}
\begin{document}
\date{\today}

\pagerange{\pageref{firstpage}--\pageref{lastpage}} \pubyear{2005}

\maketitle

\label{firstpage}

\begin{abstract}
We investigate the baryon fraction in dark matter haloes formed in
non-radiative gas-dynamical simulations of the $\Lambda$CDM cosmogony.
By combining a realisation of the \textit{Millennium Simulation}
(Springel et al.) with a simulation of a smaller volume focussing
on dwarf haloes, our study spans five decades in halo mass, from
$10^{10}~h^{-1}~$M$_\odot$ to $10^{15}~h^{-1}~$M$_\odot$. We find that
the baryon fraction within the halo virial radius is typically $90\%$
of the cosmic mean, with an rms scatter of $6\%$, independently of
redshift and of halo mass down to the smallest resolved haloes. Our
results show that, contrary to the proposal of Mo et al. (2005),
pre-virialisation gravitational heating is unable to prevent the
collapse of gas within galactic and proto-galactic haloes, and confirm
the need for non-gravitational feedback in order to reduce the
efficiency of gas cooling and star formation in dwarf galaxy
haloes. Simulations including a simple photo-heating model (where a
gas temperature floor of $T_{\rm floor} = 2\times 10^4~$K is imposed
from $z=11$) confirm earlier suggestions that photoheating can only
prevent the collapse of baryons in systems with virial temperatures
$T_{200} \lesssim 2.2~T_{\rm floor} \approx 4.4\times10^4$ K
(corresponding to a virial mass of $M_{200}\sim 10^{10}~h^{-1}~
$M$_{\odot}$ and a circular velocity of $V_{200}\sim 35~$km
s$^{-1}$). Photoheating may thus help regulate the formation of dwarf
spheroidals and other galaxies at the extreme faint-end of the
luminosity function, but it cannot, on its own, reconcile the
abundance of sub-$L_\star$ galaxies with the vast number of dwarf
haloes expected in the $\Lambda$CDM cosmogony. The lack of evolution
or mass dependence seen in the baryon fraction augurs well for X-ray
cluster studies that assume a universal and non-evolving baryon
fraction to place constraints on cosmological parameters.
\end{abstract}


\section{Introduction}

In the current paradigm of cosmic evolution, the $\Lambda$CDM model,
structures grow from an initially smooth density field into a rich
network of filaments and haloes. Initially, baryons approximately
follow the collisionless dark matter, but the two components evolve
differently in non-linear regions after recombination.  In
protogalactic haloes, for example, the kinetic energy of the collapse
is thermalised in shocks by the baryons, but rapid radiative cooling losses
allow for further collapse, leading to the formation of dense gaseous
discs susceptible to swift transformation into stars
\citep{1978MNRAS.183..341W}.

The efficient gas cooling that accompanies the collapse of
protogalactic haloes at high redshift underlies one of the central
apparent conflicts between hierarchical models of structure formation
and observation. Indeed, \citet{1991ApJ...367...45C} and
\citet{1991ApJ...379...52W} highlighted that, in the absence of
additional physics, hierarchical models predict that essentially the
entire baryonic content of the Universe should have cooled at high
redshift and, presumably, turned into stars by the present epoch. This
is in strong contradiction with observations, which suggest that stars
make up less than 5\% of the baryons in the Universe \citep[for a
review, see][]{2001MNRAS.326.1228B}. To avoid this `cooling
catastrophe', models of galaxy formation frequently invoke various
astrophysical mechanisms that counteract cooling and, in some cases,
reheat cold gas.

One consequence of the cooling catastrophe is the difficulty in
reconciling the galaxy luminosity function with the $\Lambda$CDM halo
mass function. This was recognised in the pioneering work of
\citet{1978MNRAS.183..341W}, and it is now widely accepted that a
heating mechanism (usually referred to as `feedback') is required to
explain the reduced star formation efficiency required to match the
shallow faint end, as well as the sharp cut-off at the bright end, of
the galaxy luminosity function
\citep[e.g.][]{1999MNRAS.303..188K,1999MNRAS.310.1087S,2000MNRAS.319..168C,2003ApJ...599...38B,2006MNRAS.370..645B,2006MNRAS.365...11C}

The main heating mechanisms are commonly believed to be: (i)
photo-heating by the energetic photons that reionised the Universe at
high redshift, (ii) the thermal and kinetic output of evolving stars
and supernovae (SNe), and (iii) the energy released by accretion of
matter into the supermassive black holes responsible for active
galactic nuclei (AGN). The importance of each of these processes
varies as a function of halo mass. Whereas (i) is thought to curtail
star formation in extremely low-mass haloes and substructures, (ii) is
assumed to regulate the star formation history of dwarf and normal
galaxies. The role of (iii) is expected to be most relevant for the
formation of giant galaxies, typically ellipticals found predominantly
in the dense cores of galaxy clusters \citep[e.g][]{1998Natur.395A..14R}.

Of all these, mechanism (ii) is the best studied, but there is still
no consensus concerning its role in regulating star
formation. Although it is energetically possible for SNe to suppress
star formation in galactic haloes (Dekel and Silk 1986, White and
Frenk 1991), it has long been recognised that the efficiency required
for this process to be viable is uncomfortably high
\citep{2003ApJ...599...38B,2005ApJ...631...21K}. Hydrodynamical
simulations, for example, show that under normal circumstances SNe are
far less efficient at expelling mass and reheating the IGM than
required by galaxy formation models in order to match the faint-end of
the luminosity function
\citep{1999ApJ...513..142M,2000MNRAS.314..511S}. Identifying an
additional or alternative source of feedback is therefore highly
desirable.

In a recent paper, \citet[hereafter M05]{2005MNRAS.363.1155M} proposed
an elegant alternative, namely, that the collapse of large-scale
pancakes and filaments might heat the gas prior to the assembly of
low-mass haloes. M05 sketch analytic arguments in support of the idea
that this `pre-virialisation' gravitational heating might generate
enough entropy, at low redshift ($z\lesssim2$), to inhibit
substantially the accretion of gas into low-mass haloes. If efficient
enough, this mechanism may offer a simple and attractive resolution to
the cooling crisis, and a possible explanation of the form of the
faint end of the galaxy luminosity function without the need to invoke
feedback from non-gravitational sources, such as photoionisation or
SNe.

This is clearly an intriguing proposition, and we present here a suite
of cosmological gas-dynamical simulations aimed at assessing the
viability of the pre-virialisation heating hypothesis in the
$\Lambda$CDM cosmogony. Our simulations allow us to probe a vast range
of halo masses, from $10^{10}~h^{-1}~$M$_{\odot}$ to roughly
$10^{15}~h^{-1}~$M$_{\odot}$. The simulations assume that baryons
evolve as a non-radiative fluid; this is a conservative assumption
when testing the pre-virialisation hypothesis, as radiative losses
would only serve to facilitate the collapse of baryons into
protogalactic haloes.

We also use the same suite of simulations to explore the baryon
fraction at the opposite end of the halo mass function, i.e., in
galaxy cluster haloes. Adopting a non-radiative gas approach is a reasonable
simplification here, since the majority of the intracluster medium
(ICM) has a cooling time that exceeds the age of the Universe. The
most massive galaxy clusters are of particular cosmological interest
since their baryon fractions are expected to trace accurately the
cosmic mean. Indeed, the comparison of cluster baryon fractions ($f_b$)
with the baryon density parameter ($\Omega_b$) implied by Big-Bang
nucleosynthesis calculations provides decisive evidence for a Universe
with (dark) matter density well below the critical density for closure
\citep{1993Natur.366..429W}. 

In addition, the apparent redshift dependence of cluster baryon
fractions can be used to constrain the geometry of the Universe
\citep{1996PASJ...48L.119S}, its deceleration
\citep{1997NewA....2..309P}, and by extension, the dark energy
equation of state \citep{2004MNRAS.353..457A}. These tests exploit the
redshift dependence of angular diameter distances and rely on cluster
baryon fractions being roughly universal and non-evolving over the
redshift range where they can be observed (typically $z<1$).

Our simulation suite provides the largest sample of haloes with
non-radiative gas dynamics reported to date. This allows us to
investigate the mass dependence, evolution and dispersion of halo
baryon fractions with unprecedented statistical reliability. These
results can be used to test critically the viability of the
pre-virialisation heating hypothesis and to examine the stability and
evolution of cluster baryon fractions.

In the following section we describe our simulation suite, and in
Section 3 we outline our analysis methods and main results. We discuss
them in Section 4 and conclude with a brief summary in Section 5.

\begin{table*}
\begin{center}
\begin{tabular}{rrrrrrrrr}
\hline
\hline
& $m_{\rm gas}$ & $m_{\rm dm}$ & $N_{\rm p}$ & $\epsilon_{\rm com}$ & $\Omega_{\rm b}$ & $\Omega_0$ & $L$ & $z_{\rm init}$ \\ 
& [$h^{-1}~$M$_\odot$] & [$h^{-1} M_\odot$] & & [$h^{-1}$~kpc] & & & [$h^{-1}$~Mpc] \\
\hline
\textsc{high-mass} & $3.12\times10^9$ & $1.42\times10^{10}$ & $5.0\times10^8$ & $100.0^*$ & 0.045 & 0.25 & 500 & 49  \\
\textsc{low-mass}  & $1.65\times10^6$ & $8.70\times10^6$    & $2.6\times10^6$ & 5.0   & 0.040 & 0.25 & 100     & 127 \\
\textsc{dwarf}     & $1.20\times10^4$ & $7.80\times10^4$    & $3.4\times10^5$ & 0.4   & 0.040 & 0.30 & 35.325  & 74  \\
\textsc{pancake}   & $1.20\times10^6$ & $7.83\times10^6$    & $3.5\times10^5$ & 10.0  & 0.040 & 0.30 & N/A     & 145 \\
\hline
\end{tabular}
\end{center}
\caption{Parameters for the various simulations. Note that each
simulation contained an equal number, $N_{\rm p}$, of both gas and
high-resolution dark matter particles. Softening lengths, $\epsilon$,
are quoted in comoving coordinates hence we mark the
\textsc{high-mass} simulation with an asterisk to denote that its
softening was switched to physical units at $z=3$. The
\textsc{pancake} simulation has no boxsize, $L$, since it was run with
vacuum boundary conditions.}
\label{tab:params}
\end{table*}

\section{Simulation Details}

To maximise the dynamic range of our halo sample we analyse two
simulations of different volumes. One (labelled \textsc{low-mass}) is
a high-resolution `zoomed-in' simulation of a relatively small volume
designed to study low-mass haloes. The other (\textsc{high-mass}) is
a gas-dynamical realisation of the \textit{Millennium Simulation}
\citep{Springel_short}, a large volume containing many well resolved
galaxy clusters. We address numerical convergence issues by simulating
the collapse of a `pancake' at varying numerical resolution. And,
finally, we test explicitly the robustness of our results in the
\textsc{low-mass} simulation by re-simulating at much higher
resolution a single (\textsc{dwarf}) halo with virial mass
$10^{10}~h^{-1}~$M$_{\odot}$.  The numerical parameters and other
details of the simulations are listed in Table~\ref{tab:params}.

We evolve our initial conditions in all cases using the publicly
available parallel code GADGET-2 \citep{2005MNRAS.364.1105S}, with
non-radiative\footnote{We avoid the common practice of terming
non-radiative simulations `adiabatic', since they include
non-adiabatic shocks.} baryon physics implemented by smoothed particle
hydrodynamics (SPH).

\subsection{\textsc{high-mass} simulation}

Our sample of high-mass haloes is drawn from a non-radiative
gas-dynamical realisation of the \textit{Millennium Simulation}. It
adopts the same displacement field and cosmology as the original
simulation, but has lower mass resolution: $5\times 10^8$ gas and dark
matter particles within a periodic simulation box of side
$500~h^{-1}~$Mpc. The baryon density parameter $\Omega_{\rm b} =
0.045$ results in particle masses of $m_{\rm
gas}=3.12\times10^9~h^{-1}~$M$_\odot$; for the dark matter we adopt
$\Omega_{\rm dm}=0.25$, implying a particle mass of $m_{\rm
dm}=1.42\times10^{10}~h^{-1}~$M$_\odot$. We choose a comoving
gravitational softening of $100~h^{-1}~$kpc until $z=3$, at which
point it is fixed in physical units to $25~h^{-1}~$kpc. Further
details and analysis of this simulation shall be presented in a
forthcoming paper (Pearce et al., \textit{in prep}).

\subsection{\textsc{low-mass} simulation}

In order to analyse a representative region of the Universe whilst
still resolving low mass haloes, we simulate the evolution of a
spherical region of radius $7~h^{-1}~$Mpc, identified at $z=0$ in a
simulation of a box $100~h^{-1}~$Mpc on a side with the same
cosmological parameters as the \textit{Millennium Simulation}. The
sphere was chosen at random from a sample of spherical regions with
mean density within 10\% of the cosmic mean, and devoid of haloes with
mass $M_{200}>10^{13}~h^{-1}~$M$_\odot$, in order to prevent the
region from being dominated by a single halo. Randomly placed spheres
satisfy the density selection criterion slightly less than 10\% of the
time, since the volume is dominated by underdense
regions. Approximately one-third of spheres satisfying the density
criterion lack haloes more massive than $10^{13}~h^{-1}~$M$_\odot$.

Our initial conditions are generated by resampling the region with a
greater number of particles and adding additional short wavelength
perturbations, whilst coarse sampling the external mass distribution
with multi-mass collisionless particles to reproduce the large-scale
gravitational field. Our resampling algorithm is based upon the
procedure outlined by \citet{1996ApJ...472..460F}, and is described in
detail by \citet{2003MNRAS.338...14P}. Gas is added to the high-resolution
region by splitting each particle into a dark matter particle and a
gas particle, with mass ratio given by the adopted baryon and dark
matter density parameters. This implies that each gas particle has a
corresponding dark matter `partner' associated with a unique volume
element at high redshift, a useful feature when tracing the
differences in the evolution of the two components in the non-linear
regime.

The \textsc{low-mass} simulation features $2.5\times10^6$ gas
particles of masses $m_{\rm gas} = 1.65\times10^6~h^{-1}~$M$_\odot$
and an equal number of high resolution dark matter particles, of mass
$m_{\rm dm} = 8.70\times10^6~h^{-1}~$M$_\odot$. At this resolution,
the simulated volume yields a sample of $\sim1300$ well resolved
(i.e., $N_{\rm dm} > 150$) low mass haloes at $z=0$ whilst remaining
relatively computationally inexpensive. 

\subsection{\textsc{dwarf} simulation}

In order to assess the robustness of our results for low-mass haloes
we re-simulate a single dwarf halo ($10^{10}~h^{-1} M_{\odot}$) at
much higher resolution than its counterparts in
\textsc{low-mass}. Because pre-virialisation heating is expected to be
most effective in haloes assembling late, we select for resimulation a
dwarf halo with relatively late formation time for its mass. The most
massive progenitor first exceeds half the final mass of the halo at
$z=0.6$, whilst the extensions to the Press-Schechter theory
\citep{1974ApJ...187..425P} described by \citet{1993MNRAS.262..627L}
suggest that the most probable formation time for a halo of this mass
is $z\sim2$.

We selected the halo from a parent simulation of box length
$35.325~h^{-1}~$Mpc and density parameters $(\Omega_{\rm
m},\Omega_\Lambda=0.3,0.7)$. We apply the same resimulation technique
as for \textsc{low-mass}, using $3.4\times10^5$ gas and
high-resolution dark matter particles. We adopt a baryon density
parameter of $\Omega_{\rm b} = 0.04$, which implies particle masses of
$m_{\rm gas} = 1.20 \times 10^4~h^{-1}~$M$_\odot$ and $m_{\rm dm} =
7.8 \times 10^4~h^{-1}~$M$_\odot$. At $z=0$, the halo has mass
$M_{200}=9.5\times10^9~h^{-1}~$M$_\odot$, virial radius $r_{200} =
34.5~h^{-1}~$kpc, and circular velocity $V_{200} = 35~$km~s$^{-1}$.

\begin{figure*}
  \includegraphics[width=\textwidth]{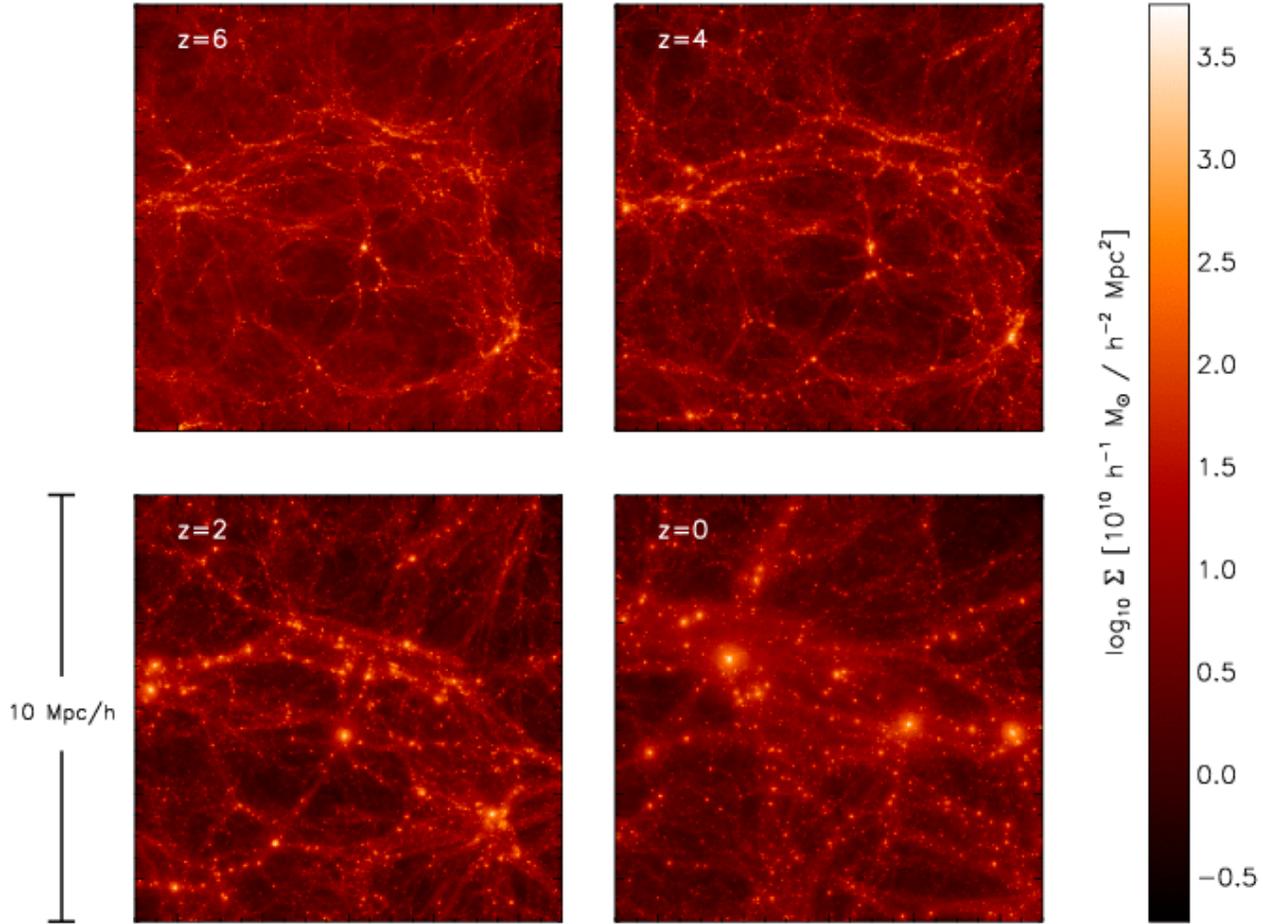}
  \caption{Redshift-progression of the projected gas density within a
  cube of side $10~h^{-1}~$Mpc (comoving) in the \textsc{low-mass}
  simulation. The progression clearly illustrates that by $z=2$ many
  dark matter haloes have already collapsed, driving their associated
  gas to high overdensities. Such gas is afforded stability against
  the shocks that develop as the large-scale environment, within which
  it is embedded, collapses into pancakes and filaments}
  \label{figs:image}
\end{figure*}

\subsection{\textsc{pancake} simulation}

The pre-virialisation mechanism outlined by M05 assumes
that gas is heated during the pancake-like collapse of the large-scale
structure within which dwarf haloes are embedded. It is therefore
important to explore whether our numerical techniques are suitable to
describe this process accurately, as well as whether the results are
not artificially marred by limited numerical resolution.

We investigate this by simulating the collapse of an idealised pancake
with similar numerical resolution as that of the \textsc{low-mass}
simulation, and then vary the resolution in order to assess
convergence.  The simulation involves the collapse of a uniform
spherical region of mass $3 \times 10^{12}~h^{-1}~$M$_\odot$, initially
perturbed so that it collapses along one axis to form a flattened
pancake at $z=2$. We use a `glass' rather than a grid in order to
minimise the artifacts introduced by anisotropies in the grid. The
desired dynamics are achieved by compressing the sphere along one axis
and expanding it by half of the compression factor along the other two
axes, with initial velocities computed using linear theory.

The choice of parameters for this simulation is motivated by the
discussion of M05, who argue that such `pancakes' might represent
the typical environment where dwarf galaxy halos are formed, and that
pre-virialisation heating might have been missed in early simulations
because of inadequate resolution.  As noted by M05, a collapsing
pancake forms shocks on both sides, so a faithful treatment of its
thermodynamic evolution requires $\gtrsim 8$ smoothing lengths across
its collapsing axis. Following M05, and assuming that shocks operate
as the collapsed structure approaches an overdensity of $\sim 10$,
this implies a pancake thickness of $200~h^{-1}~$ kpc (comoving) and a
minimum resolution of $25~h^{-1}~$kpc, again comoving. We fulfil this
criterion using a particle mass similar to that of the
\textsc{low-mass} simulation. At an overdensity of $10~\Omega_{\rm
b}\rho_{\rm c}(z)$ a mass of $40~m_{\rm p}$ particles (comparable to
what is used to define the SPH smoothing length scale) is contained
within a sphere of comoving smoothing length $h_{\rm sml}\sim
25~h^{-1}~$kpc.

\section{Analysis \& Results}

Figure~\ref{figs:image} shows various snapshots of the
\textsc{low-mass} simulation. Each box is $10~h^{-1}~$Mpc on a side
and shows the positions of gas particles, colour-coded according to
density. This figure shows clearly the highly anisotropic nature of
the large-scale structure, but also highlights the points that many
dwarf dark matter halos had already collapsed by $z\sim 2$. As we
discuss below, this has important implications for the efficiency of
pre-virialisation heating and its effects on the baryon fraction of
collapsed systems.

\begin{figure*}
  \includegraphics[width=150mm]{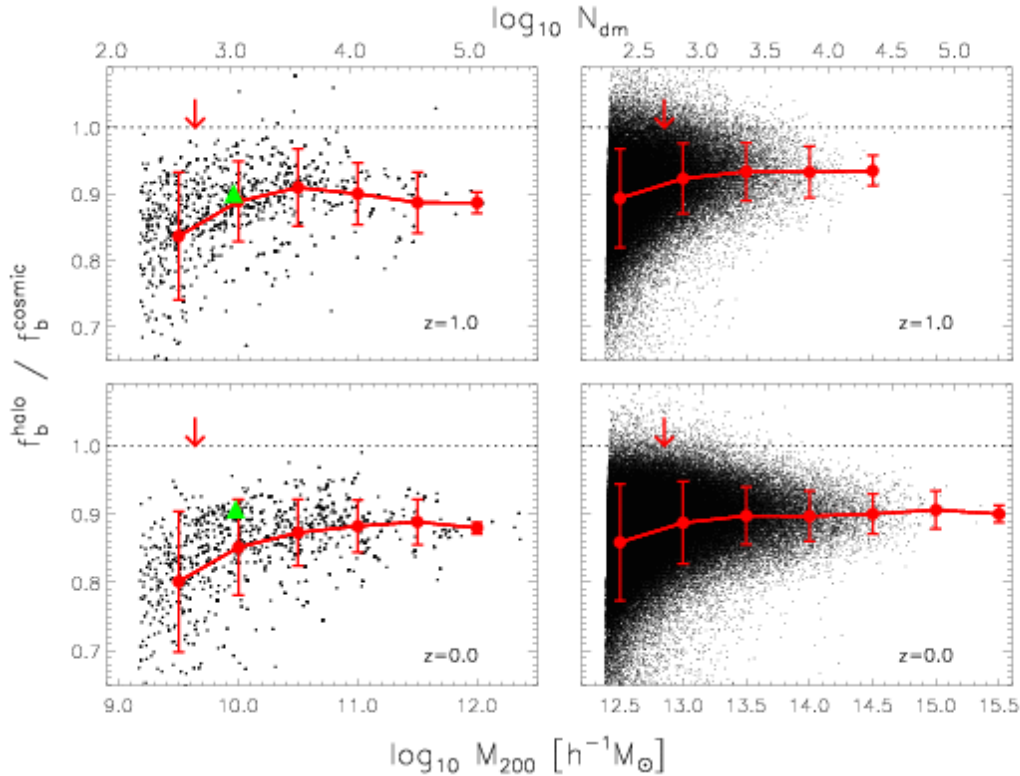}
  \caption{Baryon fractions, in units of the universal value, of well
  resolved haloes ($N_{\rm dm} > 150$) drawn from the
  \textsc{low-mass} (\textit{left-hand panels}) and \textsc{high-mass}
  (\textit{right-hand panels}) simulations at $z=1$ (\textit{upper
  panels}) and $z=0$ (\textit{lower panels}). The large dots and error
  bars show the mean and rms of the distribution, respectively. The
  green triangle marks the baryon fraction of the halo in the
  \textsc{dwarf} simulation. The upper horizontal axis gives the
  equivalent number of dark matter particles at a given mass
  scale. The downward arrows in each plot illustrate the mass scale
  corresponding to 500 dark matter particles.} \label{figs:bf}
\end{figure*}

\subsection{Halo finding algorithm}

We use a friends-of-friends group finding algorithm
\citep{1985ApJ...292..371D} with a short linking length ($b=0.05$) to
locate the cores of haloes in the simulation volumes. The centres of
these haloes are used as starting points for an iterative algorithm that
finds spherically overdense regions of mean enclosed density
$200\rho_{\rm crit}(z)$, whose centres coincide with their centres of
mass. We define the virial radius, $r_{200}$, as the radius of this
sphere. Other `virial' quantities quoted for haloes refer to
measurements within this radius, unless otherwise specified. 

We consider only haloes with at least $150$ dark matter particles in
order to minimise the effects of poor numerical resolution, and clean
the sample by removing haloes partially contained within other haloes. In the
case of \textsc{low-mass} we also disregard haloes located closer than
$200~h^{-1}$ kpc from the boundary of the resimulated region at
$z=0$, since they may have been subject to boundary effects.

\subsection{Baryon fractions}

\subsubsection{\textsc{high-mass}}

The right-hand panels of Figure~\ref{figs:bf} show the baryon fraction in haloes
identified in the \textsc{high-mass} simulation at $z=1$ and at
$z=0$. These panels show the largest sample of haloes simulated with
non-radiative gas physics reported to date, with approximately
$49,500$ and $115,000$ haloes resolved at $z=1$ and $z=0$
respectively. The panels show that the baryon fraction is independent
of mass, and has not evolved at least since $z=1$, the highest
redshift for which cluster baryon fractions can be estimated reliably
from X-ray observations (e.g. the observations of
\citet{2004MNRAS.353..457A} span the redshift range $0.07<z<0.9$).

The mean cluster baryon fraction within the virial radius is
approximately 90\% of the cosmic mean, with relatively small scatter;
the root-mean-square dispersion is less than 3\% for haloes of mass
$M_{200} \gtrsim 3 \times 10^{14}~h^{-1}~$M$_\odot$, and the
difference between the 10$^{\rm th}$ and 90$^{\rm th}$ percentiles is
always less than 7.5\% over the same mass range. The scatter remains
small for all well resolved haloes; the rms scatter is less than $6\%$
for haloes resolved by at least 500 dark matter particles. This result
is in broad agreement with previous simulations of cluster baryon
fractions in the non-radiative regime using SPH codes
\citep[e.g.][]{1995MNRAS.275..720N,1998ApJ...503..569E,Frenk_short,2006MNRAS.365.1021E}.

These results are only weakly dependent on radius; within a radius
encompassing a mean inner density 500 times greater than critical,
$r_{500}$, the results are very similar to those within $r_{200}$: for
$M_{200} \gtrsim 3\times 10^{14}~h^{-1}~$M$_\odot$ the baryon fraction
remains, on average, $90\%$ with an rms dispersion of $3\%$. This radius,
which is $\sim0.7 r_{200}$ for an NFW profile \citep{1996ApJ...462..563N,1997ApJ...490..493N} with concentration
$c=5$, is roughly the maximum radius for which total cluster masses
can be estimated reliably from X-ray observations
\citep[e.g.][]{2006ApJ...640..691V}.

\subsubsection{\textsc{low-mass}}

The left-hand panels of Figure~\ref{figs:bf} show that a similar
result applies to the halo sample identified in the \textsc{low-mass}
simulation. The baryon fraction of galactic and dwarf haloes is also
about $90\%$ of the cosmic mean, and shows no discernible dependence on
mass or redshift.

A noticeable downturn is observed below about $10^{10}~h^{-1}
M_{\odot}$, but this may be ascribed to the poorer resolution
affecting such haloes, since an underestimate of the gas density at
accretion shocks leads to artificially high post-shock entropies. Note
that a similar downturn is also seen in the \textsc{high-mass} sample
for haloes resolved with fewer than $\sim 500$ particles. Downward
arrows show the mass scale corresponding to $500$ dark matter
particles in each simulation.

\subsubsection{\textsc{dwarf}}

Figure~\ref{figs:bf} suggests that for haloes resolved with more than
$\sim 500-1000$ particles, numerical resolution does not affect
the measured baryon fractions. Further supporting evidence
comes from the results obtained for the high resolution individual
\textsc{dwarf} halo. These are shown in Figure~\ref{figs:bf} with a
filled triangle, and are consistent with \textsc{low-mass} haloes of
similar mass.

\subsubsection{\textsc{pancake}}

One may still worry that, despite the apparent convergence of the
results shown in Figure~\ref{figs:bf}, the numerical resolution is
insufficient to capture the shocks during the pancake collapse phase
that accompanies the formation of haloes. We can test this by examining
the \textsc{pancake} simulation.

\begin{figure}
\includegraphics[angle=0,width=80mm]{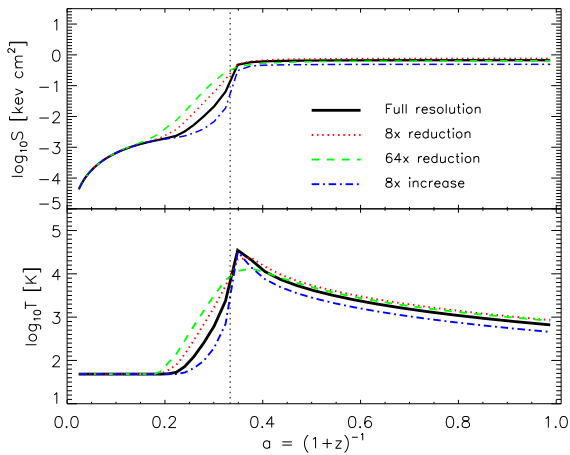}
\caption{The evolution of the median entropy parameter (\textit{top})
 and median temperature (\textit{bottom}) of gas particles within the
 \textsc{pancake} collapse simulation. The solid line in each case
 represents the simulation featuring particle mass resolution
 identical to the \textsc{low-mass} simulation. We supplement this run
 with degraded resolution runs with particle number reduced by a
 factor of 8 (\textit{dotted line}) and 64 (\textit{dashed line}). An
 improved resolution run is also presented, with particle number a
 factor of 8 greater (\textit{dot-dashed line}). The vertical dotted
 line marks $z=2$.} \label{figs:pnckent}
\end{figure}

Figure~\ref{figs:pnckent} shows the evolution of the entropy (measured
by $s=T/n_{\rm e}^{2/3}$), and of the temperature of the gas during
the collapse of the sphere to a pancake configuration. Note that when
computing the electron density, $n_{\rm e}$, we assume the gas is
fully ionised and of primordial composition. At $z=2$ the system
reaches maximum asphericity, and is well described by a plane of
comoving thickness $25~h^{-1}~$kpc (c.f. comoving softening of
$10~h^{-1}~$kpc) and comoving radial extension of $1.5~h^{-1}~$Mpc.

The collapse heats gas to a median temperature of $3 \times 10^{4}~$K,
and the median specific entropy reaches $s\simeq1~$keV~cm$^2$. We test
explicitly the effect of resolution by re-running the \textsc{pancake}
simulation with particle numbers reduced by factors of $8$ and $64$
relative to the standard resolution of \textsc{low-mass}; we complete
the resolution study by running a higher-resolution case where
the number of particles was increased by a factor of
$8$. Gravitational softenings were scaled as $\epsilon \propto N_{\rm p}^{1/3}$.

As shown in Figure~\ref{figs:pnckent} the post-shock median entropy
and temperature jumps are quite insensitive to numerical resolution,
although the transition becomes noticeably sharper as the resolution
increases. This test shows that limited resolution does not lead to a
substantial underestimate of the entropy jump in numerical simulations
of pancake-like collapse. Actually, poor resolution leads typically to
underestimation of the true densities: entropies are therefore
typically {\it overestimated} in poor resolution simulations. This
result, together with the consistency between the \textsc{low-mass}
and \textsc{dwarf} simulations, gives us confidence that baryon
fractions in our simulations are not unduly affected by resolution
effects.
 
\begin{figure}
\includegraphics[angle=0,width=80mm]{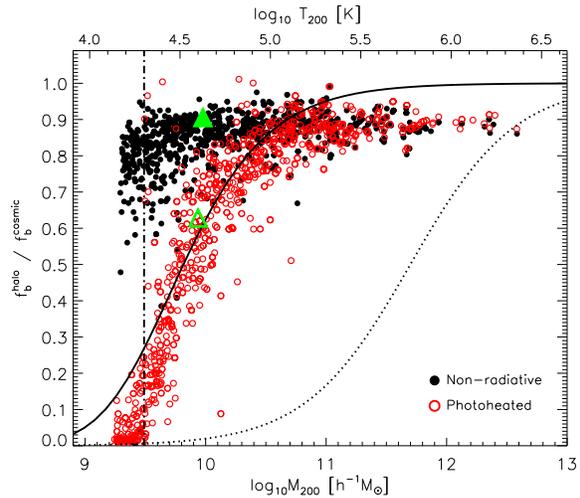}
\caption{Baryon fractions, in units of the universal value, at $z=0$
for the \textsc{low-mass} simulation with and without our
photoionising background model. The open and solid green triangles
show the baryon fraction of the \textsc{dwarf} halo with and without
the same model, whilst the vertical dot-dashed line marks the virial
mass and virial temperature equating to the temperature floor. The
solid line denotes the baryon fractions predicted by M05 for haloes in
the presence of a UV background, whilst the dotted line shows the
prediction of the M05 gravitational preheating model.}
\label{figs:fb_m05_ph}
\end{figure}

\subsection{Photoheating}

The one feedback mechanism that is certainly present at early times
is associated with the energetic photons that reionised the Universe at high
redshift. This has long been recognised as having the potential to
inhibit the formation of galaxies in low-mass haloes, although there
is still no consensus concerning the mass scale below which
photoheating becomes effective at halting galaxy formation
\citep{1984Natur.311..517B,1992MNRAS.256P..43E,1996MNRAS.278L..49Q,2000ApJ...539..517B,2002MNRAS.333..156B}. Most semianalytic galaxy formation models have so far
adopted the prescription of \citet{2000ApJ...542..535G} to determine
the gas accreted by haloes for given IGM pressure, but recent results
presented by \citet{2006MNRAS.371..401H} suggest that Gnedin's
approach may substantially overestimate the mass scale of
photoheating.

This motivates us to include a simple photionisation heating model in
our simulations. The aim is twofold. On the one hand, we wish to shed
light on the disagreement about the effects of photoheating on gas
fractions, but, on the other hand, we would also like to explore whether photoheating may
act to suppress the efficiency of gas accretion in the early phases of
the hierarchy, facilitating and enhancing the thermodynamic effect of
pancake-driven shocks.

We investigate the combined effect of gravitational and photoheating
by re-running the \textsc{dwarf} and \textsc{low-mass} simulations,
again with non-radiative gas physics, but, motivated by the third-year
\textit{WMAP} data \citep{Spergel_astroph}, imposing a spatially
uniform temperature floor for all gas particles at $z=11$. To aid as
much as possible the pre-virialisation generation of entropy, we adopt
for the temperature floor a rather high value, $T_{\rm floor}=2\times10^4$~K,
consistent with the maximum temperature of the IGM at mean density, as
probed by QSO absorption spectra \citep{2000MNRAS.318..817S}.

This should clearly impact gas accretion on dwarf galaxy haloes and, in
particular, our \textsc{dwarf} halo, where the virial temperature is
only $\sim 4.4\times10^4~$K at $z=0$, only a factor of 2.2 above $T_{\rm
floor}$. We therefore expect that a considerable fraction of the gas
bound to small halo progenitors at high redshift should be
photo-evaporated from this structure.

Figure~\ref{figs:fb_m05_ph} illustrates the effect of the additional
heating on the baryon fractions of the \textsc{low-mass} halo sample
(open circles) and compares them with the results of the non-radiative
run (filled circles). Photoheating introduces a well-defined mass scale below
which gas accretion is strongly suppressed. Below $M_{200}\sim
10^{10}~h^{-1}~$M$_{\odot}$ haloes are able to retain less than one
half of their share of baryons within their virial radii; the effect
is as large as $90\%$ in haloes below $3\times
10^9~h^{-1}~$M$_{\odot}$, corresponding to an effective virial
temperature very similar to the photoheating temperature floor.

\begin{figure}
\includegraphics[angle=0,width=80mm]{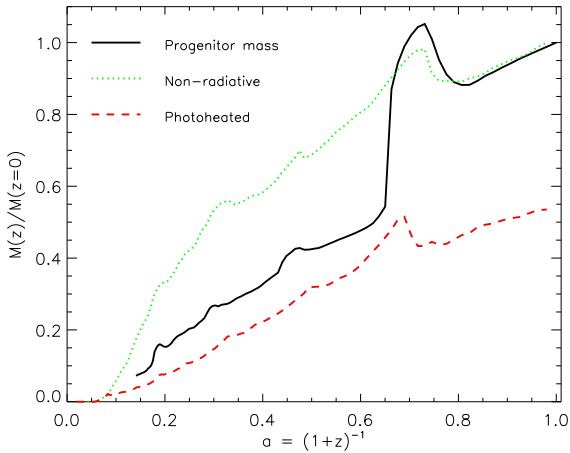}
\caption{A comparison of the evolution of the mass of the main
\textsc{dwarf} halo progenitor (\textit{black solid line}), with the
evolution of the final halo gas mass that is already collapsed
at redshift $z$, as defined by the condition that its density exceeds
$10\rho_{\rm c}(z)$ (\textit{green dotted line}). The data are
normalised to their values at $z=0$. The latter quantity is also shown
for the case when our photoheating treatment is applied (\textit{red
dashed line}), this time normalised to the $z=0$ value of the purely
non-radiative case.}
\label{figs:clmf}
\end{figure}

\section{Discussion}

The main result of the previous section is that, in the non-radiative
approximation, the baryon fraction of $\Lambda$CDM haloes is
independent of redshift as well as of mass, in the range resolved by
our simulations ($10^{10}$-$10^{15}~h^{-1}~$
M$_{\odot}$). Photoionisation reduces the baryon fraction only in haloes
with virial temperature comparable to that imposed by the ionising
photons, typically just above $10^4$ K. We discuss below the
implication of these results for models of galaxy formation and for
the interpretation of observations of baryon fractions in clusters.

\subsection{Baryon fraction bias}

The lack of dependence of baryon fractions on halo mass is intriguing,
as is the fact that the mean value within the virial radius is only
$\sim 90\%$ of the cosmic mean. The same result has been observed in
other simulations
\citep{1995MNRAS.275..720N,1998ApJ...503..569E,Frenk_short},
and has been ascribed to the collisional \textit{vs.} collisionless
nature of the baryons and dark matter, coupled to the hierarchical
assembly of haloes in the $\Lambda$CDM cosmogony. Indeed, during the
many mergers that mark the formation of a halo, shocks act to stop the
gas whilst the dark matter streams through freely. This leads to a
temporary spatial offset between dark and gaseous components during
which energy and angular momentum are transferred from the dark matter
to the baryons, as discussed in detail by
\citet{1993MNRAS.265..271N}. The energy gained during mergers results
in a more extended gaseous component, and an overall (slight)
reduction in the baryon fraction relative to the cosmic mean.

\subsection{Pre-virialisation heating}

The results of Figure~\ref{figs:bf} imply that pre-virialisation
heating is ineffective at preventing the collapse of baryons into
low-mass haloes, even for masses as low as $10^{10}~h^{-1}~$
M$_{\odot}$. Figure~\ref{figs:fb_m05_ph} compares the baryon fractions
of \textsc{low-mass} haloes at $z=0$ (filled circles) with the
predictions of the M05 model (dotted line). M05 argue that
the IGM in low-mass haloes should have been heated by shocks to
roughly $\sim 10$ kev cm$^2$ prior to halo assembly, and that this
would lead to a reduction of $\gtrsim 50\%$ in the baryons filling
haloes of mass $< 6 \times 10^{11}~h^{-1}~$M$_\odot$. M05 propose a
fitting formula to characterise this effect;

\begin{equation}
\frac{f_{\rm b}^{\rm halo}}{f_{\rm b}^{\rm cosmic}} = \frac{1}{(1+M_{\rm
c}/M)^\alpha} \label{eq:m05}
\end{equation}

where $\alpha=1$ and $M_{\rm c} = 5\times10^{11}~h^{-1}~$M$_\odot$ is
a characteristic mass scale. This function clearly fails to reproduce
our results, and suggests that the hypotheses on which M05 base their
model are not satisfied in our simulations. 

The main premise of M05's model is that most low-mass haloes form in
extremely aspherical regions where their assembly might be delayed,
allowing for pancake-driven shocks to elevate the entropy of the IGM
prior to halo assembly. Our simulations, however, indicate otherwise.

Firstly, low-mass haloes surviving to the present were, at the time of
their formation, in regions where pre-virialisation shocks were weaker
than envisaged by M05. For example, we find a typical post-collapse
entropy of $\sim 1$ keV cm$^2$ in our pancake collapse simulations,
about an order of magnitude lower than adopted by M05 to compute the
model shown by the dotted curve in Figure~\ref{figs:fb_m05_ph}. These
results suggest that the halo assembly process is approximately
scale-free; if pre-virialisation heating does indeed occur, it affects
{\it all} haloes in similar measure, leaving no particular signature in
low-mass haloes.

Secondly, the material destined to make a low-mass halo collects into
dense, early-collapsing clumps prior to the collapse of the
surrounding `pancake'. The pancake, in other words, is not a nearly
uniform aspherical structure where shocks may propagate freely, but
rather a large-scale feature where a substantial fraction of the mass
is in collapsed clumps.

We illustrate this in Figure~\ref{figs:clmf}, where we plot (dotted
line) the fraction of the final \textsc{dwarf} halo gas that resides
in collapsed structures, as quantified by the condition $\rho > 10\,
\rho_{\rm crit}(z)$. By this rather strict measure, half the
\textsc{dwarf} gas is already in collapsed structures prior to $z\sim
2$, although by then the most massive halo progenitor (solid line) has only
about $\sim 25\%$ of the final mass. This early aggregation of the
halo gas into dense structures prevents it from being shock-heated by
the pancake-driven shocks, reducing further the pre-virialisation
heating efficiency.

\subsection{Photoheating}

As shown in Figure~\ref{figs:fb_m05_ph}, the baryon fraction may be
reduced because of heating by a photoionising background, but the effect
(at $z=0$) is restricted to haloes with virial temperatures $\lesssim
2.2 T_{\rm floor}$. This implies that ionising photons, which are
unlikely to heat the gas to temperatures much higher than $\sim
2\times10^4$~K are only able to influence the formation of galaxies in
haloes less massive than $\sim 10^{10}~h^{-1}~$M$_{\odot}$.  Our
results for the baryon fraction in this case are well described by
Eqn.~\ref{eq:m05}, but with the revised parameters suggested by
M05 for their photoheating model: $\alpha=3$ and $M_{\rm c} =
1.7\times10^9~h^{-1}~$M$_\odot$; we show this fit in
Figure~\ref{figs:fb_m05_ph} as a solid line.

The results shown in Figure~\ref{figs:fb_m05_ph} agree with those of
\cite{2006MNRAS.371..401H}, who used a more detailed treatment of the
UV background; the similarity of our findings is rather
encouraging. We concur with their assessment that the characteristic
mass scale of photoevaporation is probably considerably lower than
derived from the filtering mass formalism of
\citet{2000ApJ...542..535G}. Thus, although photoheating can reduce
the baryon fraction in low-mass systems, it appears to be less
efficient at shaping the extreme faint end of the galaxy luminosity
function than previously inferred through semi-analytic modelling
\citep[e.g.][]{2002MNRAS.333..156B,2002ApJ...572L..23S}.

Our results seem to be robust to numerical resolution, as shown by the
good agreement between the baryon fraction of the \textsc{low-mass}
and the \textsc{dwarf} haloes plotted in
Figure~\ref{figs:fb_m05_ph}. In the latter case, photoheating reduces
the baryon fraction by $50\%$ but there is no evidence that
pre-virialisation has played a r\^ole; indeed, inspection of the
evolution of the collapsed gas fraction (red dashed line in
Figure~\ref{figs:clmf}) shows no discernible feature that may be
associated with the collapse of the large-scale structure.

M05's pre-virialization model, shown as a dotted line in
Figure~\ref{figs:fb_m05_ph}, requires baryon fractions to be halved in
haloes as massive as $5\times 10^{11}~$M$_\odot$; this is an order of
magnitude larger than the mass of haloes significantly affected by
photoheating and pre-virialisation in our simulations. Our results
suggest, then, that other feedback mechanisms are required to match
the faint end of the galaxy luminosity function in the $\Lambda$CDM
scenario.

\subsection{Application to cluster surveys}

Our \textsc{high-mass} simulation also features over $115,000$ galaxy
cluster-sized haloes, and demonstrates that, in the non-radiative
regime, cluster baryon fractions are independent of virial mass,
display little dispersion, and do not evolve significantly over the
redshift range $0<z<1$. Observationally, mass profiles of clusters are
typically only estimated reliably out to a maximum radius of $r_{500}$; the
results within that radius are very similar to those plotted in
Figure~\ref{figs:bf} at the virial radius. In haloes where the region
interior to $r_{500}$ is resolved by at least 500 dark matter
particles, the mean cluster baryon fraction within $r_{500}$ at $z=0$
remains approximately $90\%$ of the cosmic mean, again with an rms
dispersion of $\sim6\%$.

The key to the applicability of our results to cosmological tests is
the validity of our non-radiative treatment of the ICM. Whilst the
failure of purely non-radiative models to match some of the global
scaling relations exhibited by clusters is well documented
\citep[e.g.][]{1991ApJ...383...95E,1991ApJ...383..104K,1995MNRAS.275..720N,1998ApJ...503..569E},
this does not necessarily imply that non-radiative models give the
wrong fraction of hot baryons in observed clusters. Since the $\rho^2$
dependence of thermal bremsstrahlung implies that the X-ray emissivity of
clusters is dominated by the central region, it is possible to
obtain agreement with the observed X-ray scaling relations
by modifying only the central gas density (for instance, with radiative
cooling or AGN feedback), whilst leaving the density of the bulk of
the gas unchanged
\citep[e.g.][]{1999MNRAS.307..463B,2002ApJ...576..601V,2003ApJ...593..272V,2002ApJ...573..515M,2004ApJ...613..811M}.

It should be noted, however, that the X-ray
luminosity-temperature relation can also be explanied by models that
do affect baryons at large radii
\citep[e.g.][]{2004MNRAS.355.1091K,2005ApJ...625..588K,2006MNRAS.365.1021E}.
In such models the hot baryon fraction at $r_{500}$ can be reduced by
up to $\sim30\%$, even for clusters with $M_{200} >
10^{15}~h^{-1}~$M$_\odot$. However, such large-scale reductions by
means of cooling and star formation conflict with optical
constraints \citep[e.g.][]{2001MNRAS.326.1228B}, and the necessary
level of feedback from SNe also appears unfeasibly high \citep{2003ApJ...599...38B,2004ApJ...608...62S}.

The effect of non-gravitational processes on the large-scale
properties of rich clusters remains a source of debate. We anticipate
that analyses of the large samples of rich clusters provided by the
\textit{Chandra} and \textit{XMM-Newton} observatories will foster the
development of a definitive picture of the ICM. This picture may well
show that baryons at large radii are only minimally affected by
non-gravitational processes, thus validating our results that rich
cluster baryon fractions do not evolve for $z<1$ and exhibit little dispersion.
 
\section{Conclusions}

We have measured the baryon fractions of a large sample of haloes
drawn from a suite of non-radiative gas-dynamical simulations of the
$\Lambda$CDM cosmology. The haloes span five orders of magnitude in
virial mass, from dwarf galaxy haloes to large clusters. Within the
virial radius, the baryon fraction averages $90\%$ of the cosmic mean,
with a fairly small dispersion ($\sim6\%$ rms) and shows no dependence
on redshift for well-resolved systems. This is at odds with the
`pre-virialisation' gravitational heating proposed by
\citet{2005MNRAS.363.1155M}. Pre-virialisation, if at all present,
plays only a minor role in setting the budget of baryons that accrete
into low-mass haloes.

Photoheating, modelled here as resulting from a uniform temperature
`floor' of $2\times10^4~$K imposed on the baryons from $z=11$, is only
able to reduce the baryon fraction in haloes with virial temperatures
comparable to the photoheating floor. The absence of
a strong mass trend in the baryon fractions of low mass haloes
highlights the need for non-gravitational feedback as a means to
regulate gas cooling and star formation in low-mass haloes, in order
to reconcile the $\Lambda$CDM halo mass function with the observed
galaxy luminosity function.

At $r_{500}$, the typical maximum radius at which current X-ray
observatories can probe cluster temperatures, the baryon fraction
remains similar to that at $r_{200}$, again with similarly small
dispersion. Provided the effects of non-gravitational physics within
$r_{500}$ on rich cluster baryon fractions are small, we conclude that
measurements of the matter density parameter and the dark energy
equation of state based on cluster baryon fractions should not suffer
from large uncertainties arising from the dispersion of baryon
fractions, nor from redshift dependent systematics.

\section{Acknowledgements}
We thank Tom Theuns, Liang Gao and Alastair Edge for helpful
discussions. We extend particular gratitude to Lydia Heck at the ICC
for computing support. RAC acknowledges a PPARC studentship. VRE is a
Royal Society URF. CSF is the holder of a Royal Society Wolfson
research merit award. JFN is supported by Canada's NSERC, as well as
by the Guggenheim, Humboldt and Leverhulme Foundations.  IGM
acknowledges support from an NSERC postdoctoral fellowship. This work
was supported in part by a PPARC rolling grant awarded to the
University of Durham. The simulations described in this paper were
conducted and analysed using HPC facilities at the universities of
Durham and Nottingham.

\bibliographystyle{mn2e}
\bibliography{bibliography}
\bsp

\label{lastpage}

\end{document}